\documentclass[fleqn,usenatbib]{mnras}

\usepackage{newtxtext,newtxmath}

\usepackage[T1]{fontenc}

\DeclareRobustCommand{\VAN}[3]{#2}
\let\VANthebibliography\thebibliography
\def\thebibliography{\DeclareRobustCommand{\VAN}[3]{##3}\VANthebibliography}

\usepackage{graphicx}	
\usepackage{amsmath}	

\usepackage{amssymb}	
\usepackage{multirow}
\usepackage{threeparttable}
\usepackage{color}
\usepackage{float}

\title[Rapid FRD determination in quasi-near field]{Rapid FRD determination for multiplexed fibre systems - I. The quasi-near field model and its uncertainties}

\author[W. Sun et al.]{
Weimin Sun$^{1,2}$,
Xudong Chen$^{1}$,
Jiabin Wang$^{1}$,
Hang Jiang$^{1}$,
Anzhi Wang$^{1}$,
Qi Yan$^{1,2,4}$,
Zhenyu Ma$^{1}$\thanks{E-mail: mzy.oe@foxmail.com (Z. Ma)},
\newauthor
Shengjia Wang$^{1,2}$,
Tao Geng$^{1,2}$,
Yue Zhong$^{5}$,
Zhongquan Qu$^{5}$,
and Yunxiang Yan$^{1,2,3}$$\thanks{E-mail: yxyan@nao.cas.cn (Y. Yan)}$
\\
$^{1}$Key Lab of In-fibre Integrated Optics, Ministry Education of China, Harbin Engineering University, Harbin 150001, China \\
$^{2}$Advanced Photonics Institute, Harbin Engineering University, Harbin 150001, China \\
$^{3}$Qingdao Innovation and Development Center of Harbin Engineering University, Qingdao 266000, China \\
$^{4}$Yantai Research Institute of Harbin Engineering University, Yantai 264003, China \\
$^{5}$Yunnan Observatories, Chinese Academy of Sciences, Yunnan 650216, China \\
}

\date{Accepted XXX. Received YYY; in original form ZZZ}

\pubyear{2023}

\begin{document}
\label{firstpage}
\pagerange{\pageref{firstpage}--\pageref{lastpage}}
\maketitle

\begin{abstract}


Focal Ratio Degradation (FRD) in fibres is a crucial factor to control in astronomical instruments in order to minimize light loss. As astronomical instrumentation has advanced, the integration of large populations of fibres has become common. However, determining FRD in multiplexed fibre systems has become a challenging and time-consuming task. The Integral Field Unit for the Fiber Arrayed Solar Optical Telescope (FASOT-IFU) represents the most densely arranged fibre-based IFU in a single unit. Due to the close packing of fibres in the V-groove of the slit end, measuring FRD is particularly challenging as the output spots are prone to overlapping with adjacent fibres. In this paper, a novel method based on the quasi-near field model is proposed to enable rapid FRD measurement in highly multiplexed fibre systems like IFUs and multi-object observation systems. The principle and uncertainties associated with the method are investigated. The method's validity is demonstrated by applying it to determine the FRD in FASOT-IFU, with the achieved FRD performance meeting the acceptable requirements of FASOT-IFU, where the output focal ratio primarily falls within the range of 5.0-7.0. The results indicate that the proposed method offers several advantages, including the simultaneous and rapid measurement of FRD in multiple fibres with high accuracy (error smaller than 0.35 in $F$-ratio). Furthermore, besides FRD, the method exhibits potential for extensive measurements of throughput, scrambling, and spectral analysis.

\end{abstract}




\section{Introduction}

In modern astronomy, fibre-based astronomical instruments have found extensive use due to their flexibility in connecting the telescope's field of view to scientific instruments. Two main techniques involve the integration of fibres in astronomical instruments: multi-object observation and integral field spectroscopy (IFS) based on integral field units (IFUs). Multi-object observation uses discrete fibres to acquire the spectrum of individual celestial objects, improving observation efficiency. The number of fibres in such systems has significantly increased, ranging from hundreds to thousands, in instruments like 2dF \citep{Lewis2002The}, SDSS \citep{york2000sloan}, Subaru \citep{Sugai2015Prime}, 4MOST\citep{Jong20144MOST}, LAMOST\citep{cui2012large}, and DESI \citep{Flaugher2014The}. Multi-object surveys with fibres have greatly advanced spectral research in astronomy, pushing the astronomical research into the era of big data.

IFS employs various methods such as microlens arrays, image slicers, and optical fibres (alone or in combination with microlens arrays).  to achieve high spatial resolution and densely arranged fibres. Fibre-based IFU has the advantages of high spatial resolution with densely arranged fibres, such as the TEIFU \citep{Murray2000TEIFU}, {SAMI \citep{Croom2012The}, MaNGA \citep{Bundy2015The}, Hector IFU \citep{Bryant2011Characterization,Bryant2014Focal,wang2023hector}, and FASOT-IFU \citep{Sun2020FASOT-IFU}. Some IFUs have the new futures of adjustable operation model and replication. For example, the IFUM consists of three IFUs at three models of 'HR/STD' for excellent and standard seeing conditions, respectively, and 'LSB' for extended low surface brightness targets. VIRUS2 \citep{gary2022virus,hanshin2018virus} is a new fibre-fed IFS building on the concept of mass-replication based on the evolution of the highly replicated instrument of VIRUS \citep{Hill2008Design,Kelz2014VIRUS}. Fibre-based IFUs enable the acquisition of the three-dimensional view of entire galaxies in a single observation. IFS generates spatially-resolved spectra over a two-dimensional field, resulting in a 3D datacube. Multi-object surveys and IFS observations provide a wealth of spectral information, leading to significant advancements in astronomical research related to cosmological models, the large-scale structure of the universe, galaxy formation and evolution, and stellar physics.

Fibres offer several advantages over traditional space optics, including simpler optical structures, smaller space requirements, and improved light loss suppression compared to discrete optics setups due to fewer mirror interfaces. However, fibres also introduce light loss through focal ratio degradation (FRD). Assessing FRD is crucial for designing fibre-fed spectroscopy systems, as it helps understand system throughput and evaluate the point spread function (PSF). FRD in fibre systems is unavoidable and results in an enlarged output spot size compared to the input light after propagating through the fibre. The main causes of FRD, assuming ideal fibre conditions without material impurities or absorption, can be classified into three aspects:

i. Mode scattering: Describes the transfer of power between different guided modes, resulting in a scattering effect with a larger output angle.

ii. Geometric deformation: Bending, twisting, and inconsistencies in core size during the drawing process can cause mode exchange and worsen FRD.

iii. Stress effect: Stretching, burdening, and fibre endface polishing introduce a stress effect.

In practice, FRD is a complex composite effect. Geometric deformation, especially microbending, can also induce the stress effect, and geometric deformation and the stress effect can affect power transfer during mode exchange. Therefore, measuring FRD is a intricate task that requires effectively excluding interference in experiments. Two primary methods are used to measure FRD based on the incident condition: the collimated beam method \citep{Haynes2008Focal,Finstad2016Collimated} and the cone beam method \citep{Ramsey1988Focal,Poppett2010A,Santos2014Studying}. Both methods involve recording output spots (circular spots or rings) in a fixed position with a known distance or in several places at different distances along with the propagation direction of the central optical axis \citep{Carrasco1994A}. The collimated beam method directly demonstrates the mode scattering effect by fitting the Gaussian profile of the output ring. However, it is typically illuminated by a laser (because the laser has high brightness and accurate direction with good alignment property), which can introduce laser speckle and does not represent the practical situation of white light from celestial bodies observed with telescopes. The cone beam method, illuminated by an incoherent light source such as an LED, simulates telescope optics with or without a central obstruction. This method directly measures the input and output focal ratios and provides a clear representation of the point spread function (PSF) through fitting the profiles of the input and output spots.

However, the aforementioned methods are usually conducted for single fibre measurements at relatively large imaging distances. In the case of multi-object observation and IFS surveys, where hundreds or thousands of fibres are integrated and rearranged in slits fed into spectrographs, conducting numerous fibre tests becomes time-consuming. Additionally, the small spacing between fibres causes the output spots to overlap, making it nearly impossible to record separate spots at a distance. To address these challenges, this paper proposes a quasi-near field method for rapid FRD measurement. The principle of FRD determination of the common methods based on the output field is introduced in Section \ref{sec.principleofFRD}, followed by a detailed explanation of the quasi-near field method's principle and application conditions in Section \ref{sec.quasinearfield}. In Section \ref{sec.validationofFRD}, the validation of the FRD determination for FASOT-IFU is presented, including the testbench setup, error analysis, and preliminary measurement results, demonstrating the method's effectiveness.

%

\section{Principle of FRD determination based on output field}\label{sec.principleofFRD}

\subsection{Focal ratio in fibre system}


The focal ratio in a fibre system represents the extent of the energy distribution divergence from the fibre endface to a distant space, forming a circular spot. This concept is illustrated in Fig.\ref{fig.frdexample}. The focal ratio, denoted as $F$, can be calculated using equation(\ref{eq.focalratio}):
\begin{equation}\label{eq.focalratio}
F = \frac{f}{d}
\end{equation}
where, $f$ is the focal length, which is the distance between the fibre end and the observation position, and $d$ represents the diameter of the spot. In general, FRD causes an enlargement of the output spot size, resulting in a decrease in the focal ratio ($F$-ratio).

\begin{figure}
  \centering
  \includegraphics[width=\columnwidth]{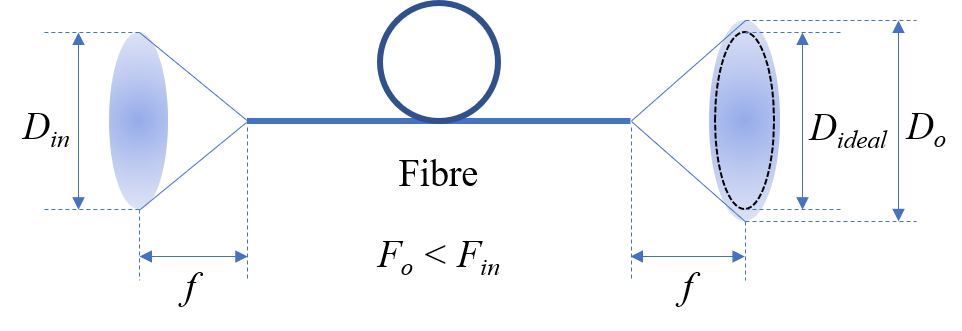}
  \caption{The concept of FRD in a fibre. Given an input $F_{in}$, the output focal ratio should be the same in the ideal situation with no FRD as presented in dashed spot. In actual operation, the output focal ratio of a smaller $F_o$ occurs, since the FRD enlarges the output spot as shown in solid line. The diameter of the output spot $D_o$ is larger than the ideal condition of $D_{ideal}$ that should be the same as the input light of spot size $D_{in}$}\label{fig.frdexample}
\end{figure}

In a telescope with a fibre-fed spectrograph system, the input focal ratio is determined by the design of the guiding optical system. It is important to ensure that the input $F$-ratio of the spectrograph is not larger than the output focal ratio of the fibre to avoid energy loss. However, achieving a small $F$-ratio, also known as a fast focal ratio, can significantly increase the complexity and cost of developing the spectrograph. Therefore, when designing a spectrograph, it is essential to consider the potential effects of FRD and provide adequate redundancy to accommodate any significant FRD in the fibre system.

\subsection{Common methods for measuring FRD}

FRD characterization methods can be categorized into imaging methods and energy encircling methods. Imaging methods involve recording the output field images at different observation planes and measuring the diameter of the spots with a certain encircled energy (EE) ratio. This can be done by calculating the spot size with respect to the total energy within the spot image frame or by measuring the full width at half maximum (FWHM) for collimated light injected at a certain angle.

One representative energy encircling method is DEEM (Direct energy encircling method) \citep{yan2018deem}, which employs a close-loop feedback system with dual power meters to directly measure the energy with a specific EE ratio. This method eliminates the need for imaging and simplifies the measurement process.

\begin{figure*}
  \centering
  \includegraphics[width=14cm]{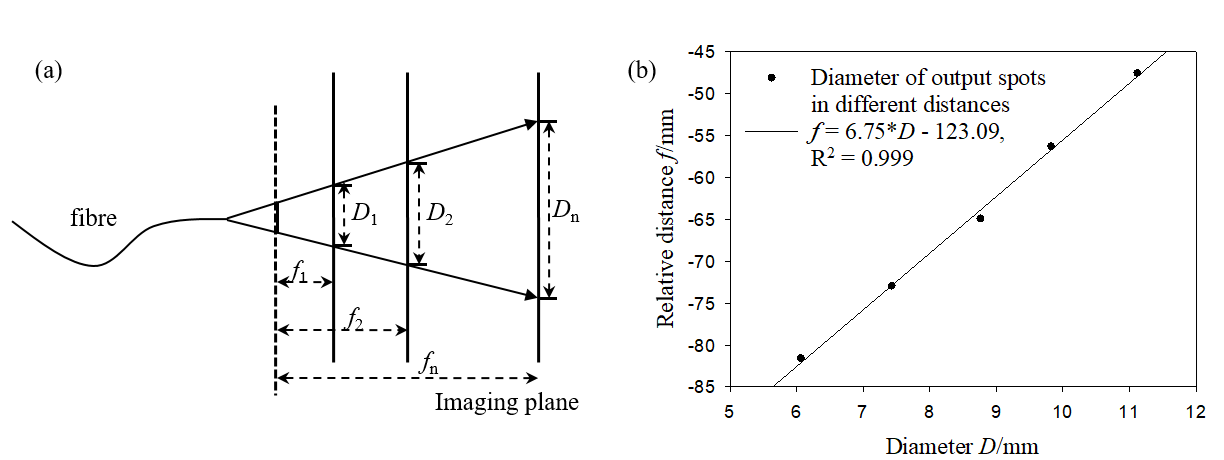}
  \caption{Common imaging method of FRD determination in far field. (a) The output spots generated by the fibre system are recorded in different imaging planes located at distances $f_n$, where $n=1,2,3,\dots$. These imaging planes are positioned at varying distances from the fibre end. (b) To determine the focal ratio of the fibre system, the diameters of the output spots are measured in each imaging plane. A linear fitting curve is then applied to the data, where the slope of the curve represents the output focal ratio. Additionally, the curve intersects the vertical axis, providing the fitted zero point of the imaging plane. This zero point serves as a reference for evaluating the accuracy of the FRD measurement. This common imaging method involves capturing the output spots at different imaging planes and analyzing their diameters to determine the focal ratio of the fibre system. The linear fitting curve enables the extraction of relevant parameters and provides insights into the accuracy of the FRD measurement.}\label{fig.frdexampled}
\end{figure*}

In these methods, the spot size is tracked at different observation planes, and it typically increases with the imaging distance. CCD cameras are positioned at multiple distances ($f$) along the normal direction of the fibre end's central axis. The spot diameter ($d$) is determined based on the chosen EE ratio, usually 90 or 95 per cent. The barycentre of the spot is first determined as the reference for spot size measurement. The encircled energy is then calculated within the entire spot image frame using the EE ratio. The focal ratio can be evaluated using Equation (\ref{eq.focalratiod}), which involves comparing the distance and spot size measurements:
\begin{equation}\label{eq.focalratiod}
F = \frac{{{f_i}}}{{{d_i}}} = \frac{{{f_j}}}{{{d_j}}} = \dots = \frac{{{f_i} - {f_j}}}{{{d_i} - {d_j}}} = \frac{{\Delta f}}{{\Delta d}}
\end{equation}
where $i$ and $j$ represent different measurements. Therefore, the ratio of corresponding distance and the diameter or the ratio of the relative difference should be always the same. Linear regression is then applied to the distance and spot size distribution to determine the slope, which corresponds to the focal ratio as shown in Fig.~\ref{fig.frdexampled}.

This method is suitable for far-field measurements where the fibre can be treated as an ideal point source, and the spot size changes linearly with the observing distance. It is commonly used for FRD measurements of individual fibres, where the camera can be placed far from the fibre endface. However, it is not applicable for multiplexed fibre systems like fibre-fed multi-object spectrographs or fibre-based IFUs. In these cases, the fibres are closely arranged, and placing the camera far away would result in overlapping spots, making individual spot identification challenging, as shown in Fig.~\ref{fig.overlapspots}. To address this, a quasi-near field method is proposed for FRD determination, specifically designed for multiplexed fibre systems. This method overcomes the difficulties associated with close-packed fibres. The details of this method and the FRD measurement of FASOT-IFU are discussed in a separate paper (Sun et al., in preparation).

\begin{figure}
  \centering
  \includegraphics[width=\columnwidth]{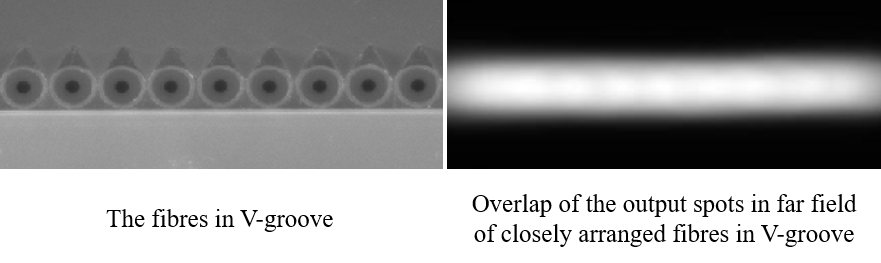}
  \caption{The figure depicts the arrangement of fibres in a V-groove and illustrates the overlapping spots in the far field. The emitted light from each fibre forms a diverging cone beam, and the narrow spacing between the fibres greatly restricts the available space for the adjacent spots to propagate separately.}\label{fig.overlapspots}
\end{figure}

\section{Quasi-near field method for FRD determination}\label{sec.quasinearfield}

\subsection{Quasi-near field model}

The output field of a fibre can be approximated as a diverging light cone. In the case of far-field measurements, where the camera is positioned several centimeters away from the fibre end, the edge of the light cone can be simplified as a linear distribution with a solid cone angle. This simplification allows for the application of ray-tracing theory and is the basis for FRD measurements using common imaging methods. Far-field measurements treat the fibre as a point source, even for large-core fibres like those with a core size of 320\,$\umu$m in LAMOST.

However, advancements in astronomical observation techniques, particularly in multi-object and integral field spectroscopy, have led to an increased number of fibres being used. To achieve higher spectral resolution, the fibre core size has become smaller, resulting in denser arrangements of fibre arrays at the slit end of spectrographs. This denser arrangement makes FRD measurements in highly-integrated multiplexed fibre systems challenging. In these cases, the CCD camera's movement range is restricted, and the imaging plane is brought closer to the fibre output end. The distance between the imaging plane and the fibre endface needs to be small enough, on the scale of millimeters, to ensure that the output spots from each fibre can be distinguished. This distance depends on factors such as the interval between the fibres in the pseudo-slit or the array end and the output focal ratio.

\begin{figure*}
  \centering
  \includegraphics[width=\hsize]{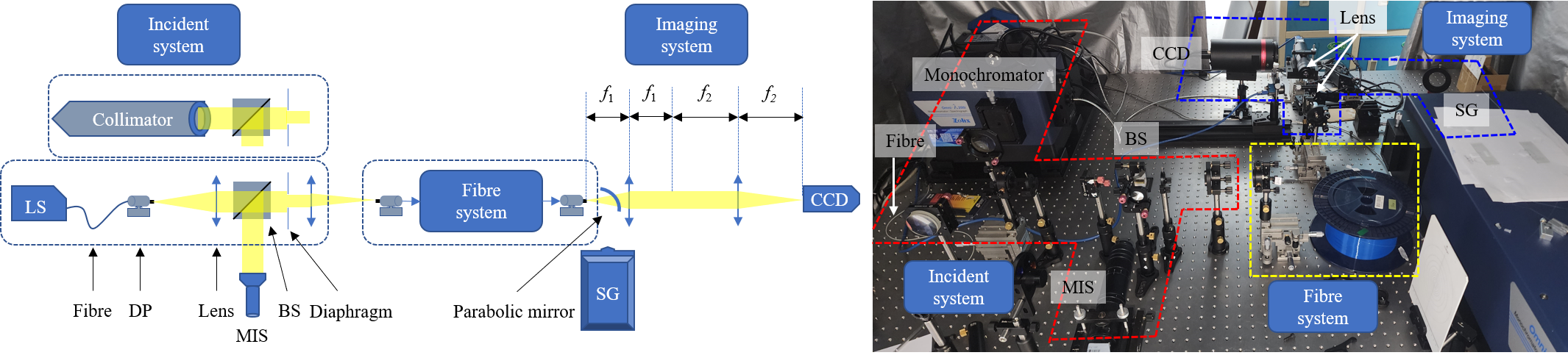}
  \caption{The figure depicts the testbench setup used for the quasi-near field method in FRD measurement with the 4F magnifying imaging system. The testbench consists of three main components: the incident system, the fibre system, and the imaging system. Additionally, a parabolic mirror is included to direct light into the spectrograph for spectral analysis. The incident system allows for precise control of the input focal ratio and facilitates inspection of the input PSF and the incident position on the fibre endface using the microscopic imaging system. In the case of an integral field unit (IFU) with a microlens array attached to the fibre array, the incident system can be replaced with a collimator to achieve similar functionality. The fibre system employed can be either a single fibre or a multiplexed fibre bundle, depending on the specific application. It serves as the conduit for transmitting light from the incident system to the imaging system. The imaging system plays a critical role in the testbench setup as it magnifies the output spots and provides an extended operational space for effectively recording the output spots at a relatively long distance. This capability ensures accurate measurement and analysis of the output spots. To facilitate spectral analysis, a parabolic mirror is incorporated to reflect the light from the fibre system into the spectrograph. The image on the right showcases the testbench setup specifically designed for conducting FRD and spectral tests on a small-scale fibre bundle containing seven fibres. Various components are labeled in the image, including the light source (LS), the six-dimensional displacement platform (DP), the microscopic imaging system (MIS), the beam splitter (BS), and the spectrograph (SG).
  }\label{fig.testbench}
\end{figure*}

For example, in the Fiber Arrayed Spectrograph for Optical Tomography Integral Field Unit (FASOT-IFU), the optical design requires an output focal ratio larger than 5.0, and the interval between fibres in the V-grooves is 130\,$\umu$m. Therefore, the imaging plane should be no further than 780\,$\umu$m at $F=6.0$ to ensure that the output spots can be separated. In such a small space, the conditions for FRD measurement significantly differ from the far-field case. The size of the fibre core cannot be ignored, and the small imaging area makes it difficult to capture enough images of the spots at different distances. Additionally, the output spot size becomes very small (less than 130\,$\umu$m for FASOT-IFU), requiring the camera's pixel size to be small enough to enhance resolution. To address these challenges, the concept the quasi-near field method for FRD measurement is proposed. It modifies the common imaging method by using a magnifying 4F imaging system to provide a feasible approach for FRD measurement. The design of the quasi-near field experiment setup is shown in Fig.~\ref{fig.testbench}.

\subsection{Quasi-near field method}

The application of this method should be considered in two situations:

i) quasi-near field with modified imaging method, and

ii) quasi-near field with parallel analogy method.

The optimal choice between these two methods can be determined based on the Fresnel number ($N_F$), which helps assess the suitability of the measurement setup.

\subsubsection{Quasi-near field with modified imaging method}

The fibre acts as a waveguide for transmitting light and also serves as an aperture when the light propagates out from the fibre end. The Fresnel number is originally introduced in the context of diffraction and describes different diffraction scenarios. When a light wave passes through an aperture and propagates over a certain distance, the diffraction phenomenon can be characterized using the Fresnel number, which is calculated using the equation:

\begin{equation}\label{eq.fn}
N_F = \frac{a^2}{L\lambda}
\end{equation}
where, $a$ represents the characteristic size of the aperture, $L$ is the distance between the aperture and the observation plane, and $\lambda$ is the wavelength. When the Fresnel number is significantly less than 1, corresponding to a relatively large distance $L$ (several centimeters), Fraunhofer diffraction occurs. In this case, the distance between the fibre end and the observation plane is sufficiently large, and the far-field measurement using the common imaging method can be employed to determine FRD. This approach is typically convenient for measuring the FRD of a single fibre.

However, in the case of FRD measurement in a multiplexed fibre system, where simultaneous measurement of the FRD for multiple fibres is desired, the CCD camera must be positioned within a few millimeters of the fibre output end to ensure that the output light spots from each fibre do not overlap with one another. At this proximity, the Fresnel number is approximately around 1. In such scenarios, the magnifying effect of the imaging system can be utilized to improve the experimental environment and expand the operating range of the CCD camera, enabling the implementation of the common imaging method to capture images at several positions. During the imaging process, the cone angle of the output field is altered by the magnifying effect, and the final focal ratio needs to be modified using the equation:

\begin{equation}\label{eq.focalratiomodify}
F = \frac{F_m}{\alpha }
\end{equation}
where, $F_m$ represents the measured focal ratio in the image space, and $\alpha$ denotes the amplification ratio of the imaging system. The amplification ratio is constrained by various factors, including the size of the camera sensor, the actual exit focal ratio of the fibre, and the aberrations of the imaging system. Ensuring that the output spots from the fibres do not overlap within the movable range of the camera is crucial in this imaging process.

\subsubsection{Quasi-near field with parallel analogy method}

When the density of the fibre array increases and the spacing between fibres becomes smaller, the non-overlapping area is significantly reduced, resulting in a Fresnel number larger than 1. In such cases, the distance between the CCD camera and the fibre ends must be much shorter to separate each spot. Even with the use of the improved conventional imaging method using the magnifying 4F imaging system, the operational space of the CCD camera remains limited. Additionally, employing the conventional imaging method in a confined space can introduce larger measurement errors due to the small changes in the fibre's output field and the limited movement distance of the CCD camera. Consequently, the conventional imaging methods are not practical for measuring FRD in these scenarios.

To address this issue, the parallel analogy method under the quasi-near field condition is proposed as an efficient approach for simultaneous FRD measurement of multiple fibres. This method is based on the fundamental definition of focal ratio. According to the linear relationship between the focal length of the fibre output focal ratio and the change in spot diameter, if the output spots from two fibres are at the same distance from the CCD camera and the direction of the optical axis of the output light cone is parallel, the output focal ratio of one fibre can be determined based on the known output focal ratio of the other fibre using the following formula:

\begin{equation}\label{eq.pam}
\frac{F_m}{F_r} = \frac{D_r}{D_m}
\end{equation}
where, $F_m$ is the focal ratio of the fibre being tested, $F_r$ is the focal ratio of a reference fiducial fibre, $D_r$ and $D_m$ represent the diameters of the spots produced by the reference fiducial fibre and the fibre being tested, respectively. The actual output focal ratio of the fibre being tested can then be calculated using equation (\ref{eq.focalratiomodify}).

The parallel analogy method offers a rapid and efficient approach for measuring FRD in multiple fibres. In this method, the fibres are densely packed in the V-groove, resulting in small distances between the output spots of adjacent fibres, as shown in Fig.~\ref{fig.pam}. By utilizing the magnifying imaging system, the distance between the fibre end and the non-overlapping spots is increased, providing a more spacious operating space for the CCD camera. Additionally, the magnifying imaging system allows the output spot to occupy more pixels on the CCD, reducing the error associated with changes in occupied pixels and improving the accuracy of spot diameter measurement, thereby enhancing the precision of FRD measurement.

During the measurement process of the parallel analogy method, the distance between the fibre end face and the CCD camera remains relatively fixed, and the change in focal ratio of each fibre is mainly determined by the size of the output spot. Therefore, in practical measurements, a suitable magnification rate $\alpha$ of the 4F system is set and maintained constant throughout the entire measurement cycle. Only the output focal ratio of the reference fiducial fibre needs to be separately calibrated. For other testing fibres, only a single spot size measurement is required, and the actual output focal ratio can be calculated based on equations (\ref{eq.focalratiomodify}) and (\ref{eq.pam}).

\begin{figure}
  \centering
  \includegraphics[width=\columnwidth]{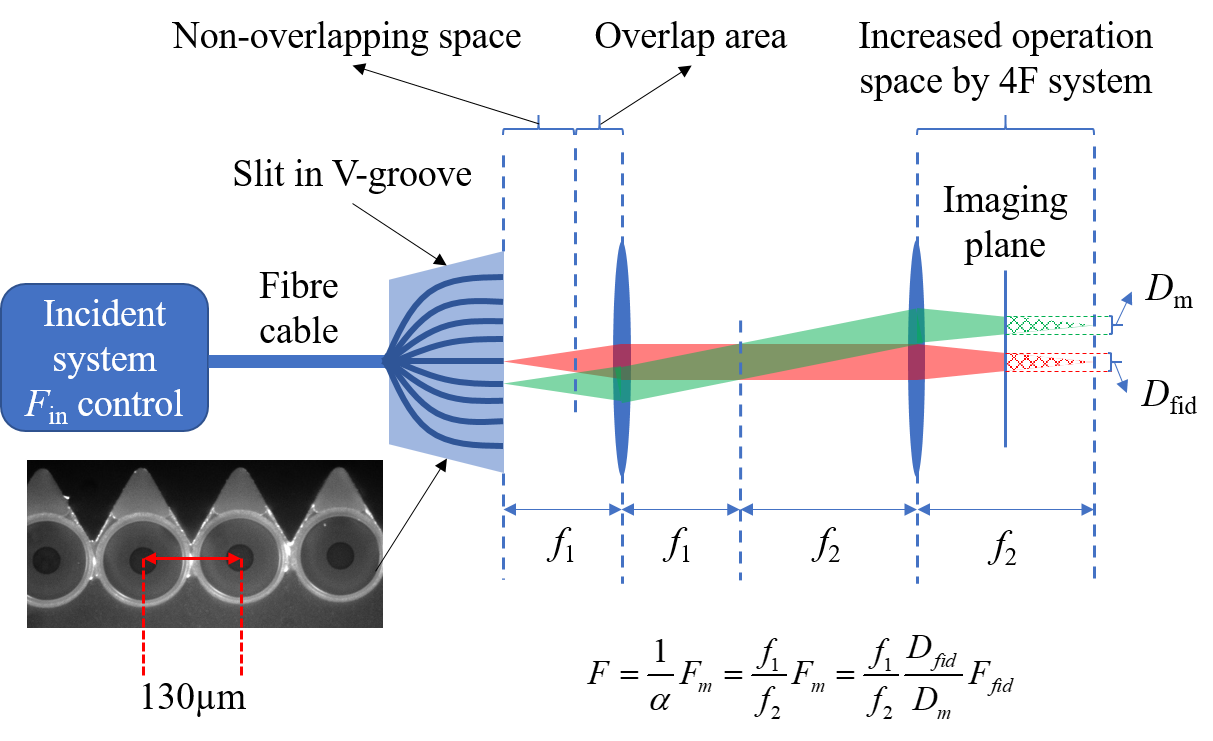}
  \caption{Quasi-near field method with parallel analogy. The 4F magnifying imaging system is utilized to expand the available operation space. In multiplexed fibre systems, densely packed fibres are positioned in a V-groove, resulting in limited non-overlapping space between adjacent fibre spots. By incorporating the 4F magnifying imaging system, the operation space can be effectively increased. Furthermore, a key advantage of this setup is that the optics preserves the propagation direction of the output spots during the transformation in the 4F system. Consequently, the focal ratio is solely determined by the magnification ratio $\alpha$, relative to the reference focal ratio of the fiducial fibre. Since the imaging plane remains unchanged, the focal ratio is exclusively influenced by the spot sizes $D_m$ of the testing fibre and $D_{fig}$ of the fiducial fibre.
  }\label{fig.pam}
\end{figure}

\section{Validation of FRD measurement for FASOT-IFU}\label{sec.validationofFRD}

The measurement of the focal ratio for a fibre is highly sensitive and prone to errors caused by various factors. In this section, we will first describe the setup of the quasi-near field method, and then explore the sources of error of the methods to minimize uncertainties. These steps are crucial for validating the measurement results. Finally, we will present the FRD measurement results obtained for the FASOT-IFU.

\subsection{Measurement setup}

FASOT is an advanced solar telescope that has been specifically designed to observe the Sun's magnetic field and photosphere \citep{Qu2011A}. FASOT achieves this through the use of an Integral Field Spectrograph (IFS) that operates at wavelengths ranging from 400 to 900\,nm, with a very high resolution of 110,000. The FASOT-IFU is a crucial component of the telescope's highly multiplexed fibre system, which contains a total of 8,064 individual fibres. The fibre structure is 35/105/125\,$\umu$m from the core to the coating layer with the NA of 0.12. The IFU is composed of two separate units, each with 4,032 fibres. These fibres are carefully arranged and separated into 12 slits that feed into the telescope's spectrographs (as shown in Fig.~\ref{fig.fasotifu}). At the input end, the IFU heads are fitted with a microlens array that is arranged in a hexagonal pattern, ensuring that there is no dead space between the fibres and providing a 100 per cent space coverage ratio. At the output end, the fibres are fixed in V-grooves to form a pseudo-slit that feeds into the spectrograph. To achieve both high spatial sampling and high spectral resolution, fibres with small core sizes are preferred. As such, the IFU employs fibres with a core size of 35\,$\umu $m.

\begin{figure}
  \centering
  \includegraphics[width=\columnwidth]{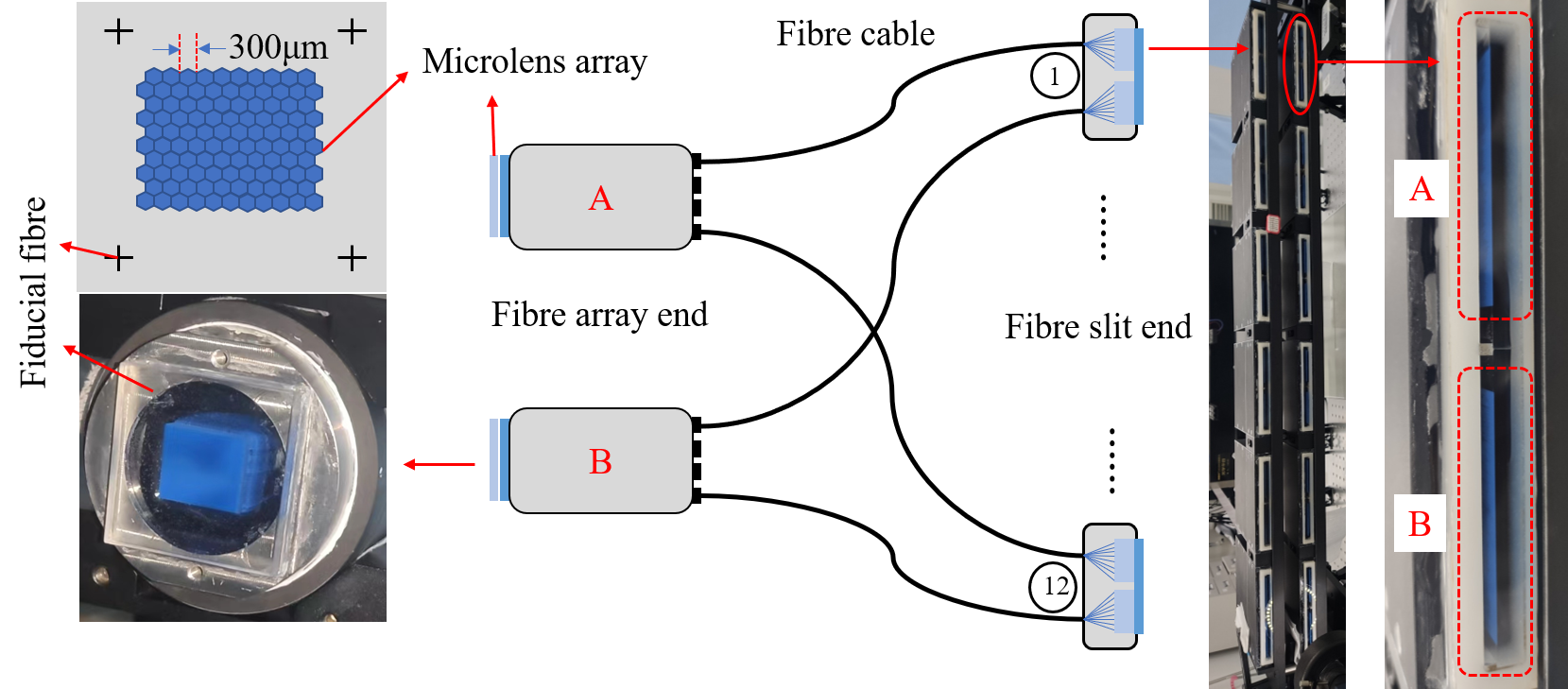}
  \caption{Fibre arrangement in FASOT-IFU. The FASOT-IFU configuration consists of two IFUs, namely A and B, each equipped with a 63$\times$64 hexagonal microlens array. Each microlens has a size of 300\,$\umu$m from edge to edge. The total number of fibres in each IFU is 4032. The fibres are organized into 12 slits, with each slit accommodating two groups of fibres, A and B. Each group comprises 336 fibres, originating from the respective IFU heads A and B. }\label{fig.fasotifu}
\end{figure}

The microlens size at the input end is 300\,$\umu$m from edge to edge, the effective focal length is 2.7 mm, and the interval between fibres in the V-grooves is 130\,$\umu$m. The compact structure makes the output spots prone to overlap with each other as \textcolor[rgb]{0.00,0.00,0.00}{shown in Fig.~\ref{fig.overlapspots}.} But measuring the FRD of each fibre individually would be a difficult and laborious task. Hence, the quasi-near field model is employed for rapid FRD measurement. The design of the FASOT-IFU requires an output focal ratio greater than 6.0, limited by the spectrograph, and an input focal ratio of $F_{in}=217$ from the guiding optical system. To prevent overlap of adjacent spots, the imaging plane should be less than 780\,$\umu$m. For simplicity, the wavelength is set to 500 nm, and the observation distance is 500\,$\umu$m, which is reasonable in case of significant FRD. Then, the Fresnel number can be estimated using Equation~(\ref{eq.fn}):
\[  N_F = \frac{a^2}{L\lambda} = \frac{17.5^2}{500 \times 0.5} = 1.225 > 1 \]

Therefore, the parallel analogy method in quasi-near field is preferred for measuring the FRD. The testbench is shown in Fig. \ref{fig.pambench}. The light source is a LED integrated in the collimator. The imaging system is fixed on the motorized multi-axis translation stage to record the output spots. The 12 slits units are parallel to the translation stage.

\begin{figure}
  \centering
  \includegraphics[width=\columnwidth]{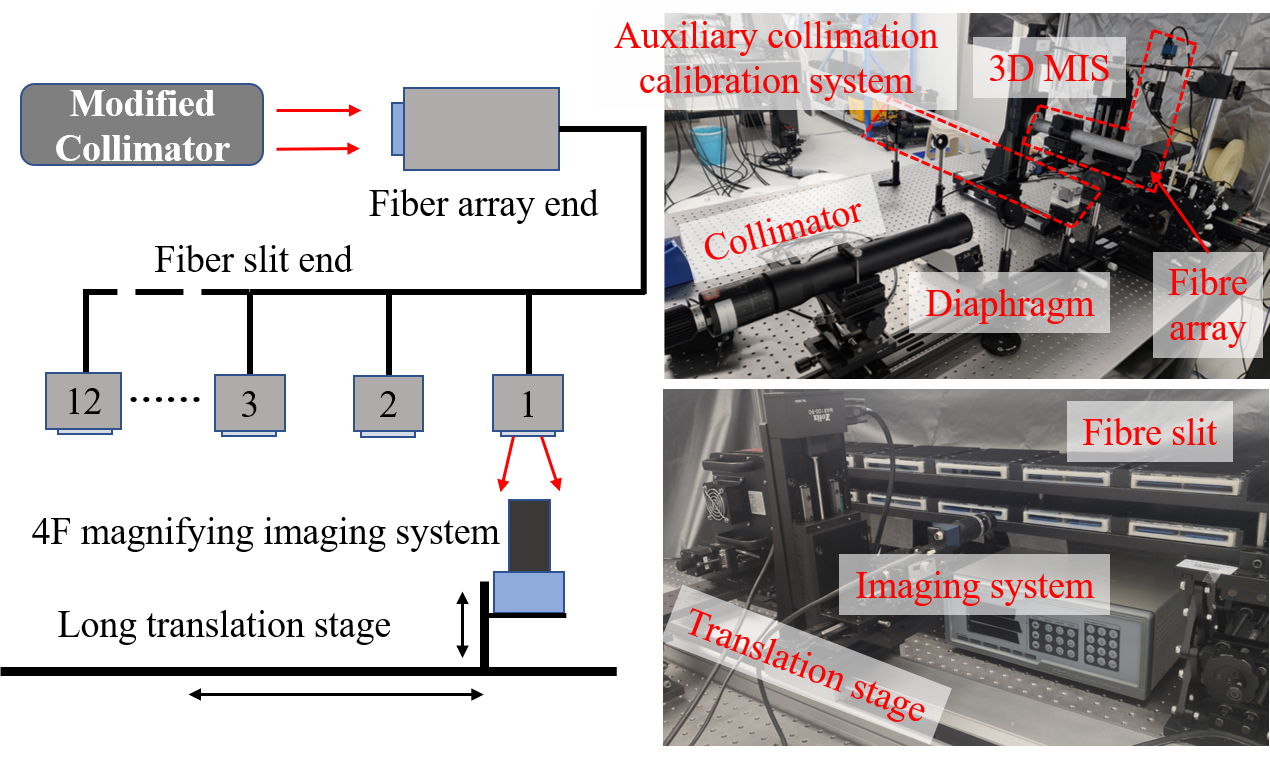}
  \caption{The diagram illustrates the quasi-near field method for FRD measurement applied to FASOT-IFU. The modified collimator is utilized to replicate the incident light from the guiding optics of FASOT, mimicking a focal ratio of 217. The imaging system can be moved parallel to the 12 slits on the long translation stage to capture the output spots. The auxiliary collimation calibration system is employed to ensure precise incident direction. The 3D microscope imaging system (3D MIS) is used for observing the fibre array. In the quasi-near field method, a single output spot per fibre is sufficient for FRD determination. These spots are then processed in the data processing pipeline to calculate the output focal ratio. The imaging system scans the spots simultaneously, significantly reducing the measurement process and saving time.
  }\label{fig.pambench}
\end{figure}


\subsection{Error analysis}


Accurately measuring the distance $f$ and spot size $D$ is crucial for estimating the focal ratio based on the principle of focal ratio. Any uncertainties that affect $f$ and $D$ will contribute to the error in the focal ratio. Misalignment in both the input and output ends of the fibre can increase the error. In addition, the computational accuracy of estimating the spot size can also introduce limitations. Furthermore, the tolerance for error varies and depends on the actual input/output focal ratio. This is because the non-linear relationship between the focal ratio and the half cone angle of the spot results in different influences on the FRD from variations in the diameters or focal distance for different focal ratios. For example, the FASOT-IFU has a focal ratio of around 6, which is much less than the cone angles of the critical numerical aperture (NA) of the fibres. The NA is defined as the sine of the maximum angle of an incident ray with respect to the fibre axis that determines the maximum acceptance angle of the incoming light. Therefore, the approximated linear relationship between the inverse of the focal ratio and the sine of the half cone angle $\theta$ is valid:
\begin{equation}\label{eq.sinna}
  sin\theta \approx \frac{D/2}{f} = \frac{1}{2F}
\end{equation}

For small angles within the limitation of NA less than 0.22, the change in sin$\theta$ is directly proportional to the variation in NA, and can be expressed as follows:
\begin{equation}\label{eq.dna}
  \Delta NA = sin(\theta _2 - \theta _1) = \Delta \theta
\end{equation}

After that, the variation in the spot size or focal distance can be converted into uncertainties in NA using the linearity relationship. This simplifies the characterization of changes in the focal ratio.

\subsubsection{Precision of the alignment}



The measurement of the focal ratio for a fibre can be influenced by several alignment issues. In this testbench, three alignment issues were considered: (a) the incident focal ratio into the fibre, (b) the offset of the fibre center to the microlens array, and (c) the parallel misalignment between the CCD and the output fibre end.

The microlens array determines the input focal ratio into the fibre, and it supports the focal ratio of $F/8$ under the illumination of parallel light. To simulate the FASOT incident condition, a collimator (focal length = 500\,mm) with a pupil of 2 mm was added in the focal point to generate quasi-parallel light with a focal ratio of $F/250$, which is very close to the guiding optical system of FASOT. The distance between the microlens array and the collimator is 1.5\,m, so error in NA caused by the incident light is smaller than 6$\times$10$^{-4}$, which is negligible for FRD measurement.

The microlens array is fixed onto the IFU head using ultraviolet curing adhesive (NOA61). Due to diffraction effects, the output spot from the microlens measures 28\,$\umu $m, which is smaller than the 35\,$\umu$m fibre core. To guarantee efficient light coupling into the fibre, the tolerance for center misalignment between the microlens array and the IFU head is limited to 5\,$\umu$m.

In the output end of the slit, the field of view of the imaging system of the camera can cover about 50 fibres in each frame. However, only 10 spots in the central area are selected for FRD measurement, since other spots are seriously distorted caused by the aberrance as shown in Fig.~\ref{fig.aberrance}. The camera should move from the first slit to the last one and the total length of the displacement is about 100\,cm, and the maximum error of the distance between the CCD camera and the fibre end in the V-groove is less than 1\,mm. Therefore, the error in NA is smaller than 0.001.

\begin{figure}
  \centering
  \includegraphics[width=\columnwidth]{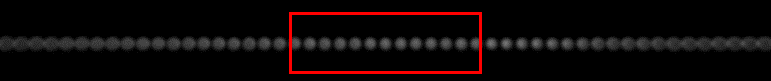}
  \caption{The figure shows the output spots in the full field of view captured by the CCD camera. Due to aberrations in the imaging system, the PSF and power distribution profiles of the output spots are distorted. As a result of these imaging limitations, only the spots within the central area (indicated by the red box) are chosen for FRD measurement. The presence of aberrations restricts the efficiency of the quasi-near field method. However, it also indicates that the method has potential for improvement by upgrading to a higher-quality imaging system capable of capturing more accurate spot profiles.}\label{fig.aberrance}
\end{figure}

\subsubsection{Uncertainties in spot size measurement}

To estimate the FRD of the output spot, it is crucial to accurately measure the diameter of the spot, which directly determines the focal ratio. Typically, the spot size is determined by measuring the 2D profile cut across the barycentre that derived from weighted average. However, due to the presence of burrs caused by fluctuations in intensity, the power distribution of the spot is not always smooth along the 2D profile cut. To avoid this issue and minimize accidental errors, a 3D fitting method is used to measure the diameter \citep[see][figure 7]{yan2018deem}). This method takes into account the data of the entire pixels within the output spot image, rather than just a cut or barycentre, thus providing a more accurate and reliable measurement of the diameter.

\subsubsection{Error dependency on focal ratio}


Equation~(\ref{eq.sinna}) shows that the relationship between the half cone angle and the focal ratio is nonlinear. In a slow input focal ratio condition, even a small change in the half cone angle can result in significant variations, whereas in a fast focal ratio case, the impact is relatively minor. Consequently, the error is dependent on the actual focal ratio. Since the output focal ratio of the FASOT-IFU is around $F/6$, it is crucial to restrict the maximum error tolerance to achieve precise estimation of the FRD performance.

The parallel analogy method involves three steps. Firstly, other testing fibres are obstructed, and the output focal ratio of the fiducial fibre is measured in the far field using common imaging methods as a reference. Secondly, the output spots array of the fibre slit, including the fiducial fibre, is recorded in the quasi-near field, and their diameters are estimated. Finally, the focal ratios and error tolerances are determined. During the imaging process, it is important to note that the fibre can be considered as an ideal point source in the far field of the fiducial fibre, but a finite extended source in the case of quasi-near field, as shown in Fig.~\ref{fig.nearandfarfield}.

\begin{figure}
  \centering
  \includegraphics[width=\columnwidth]{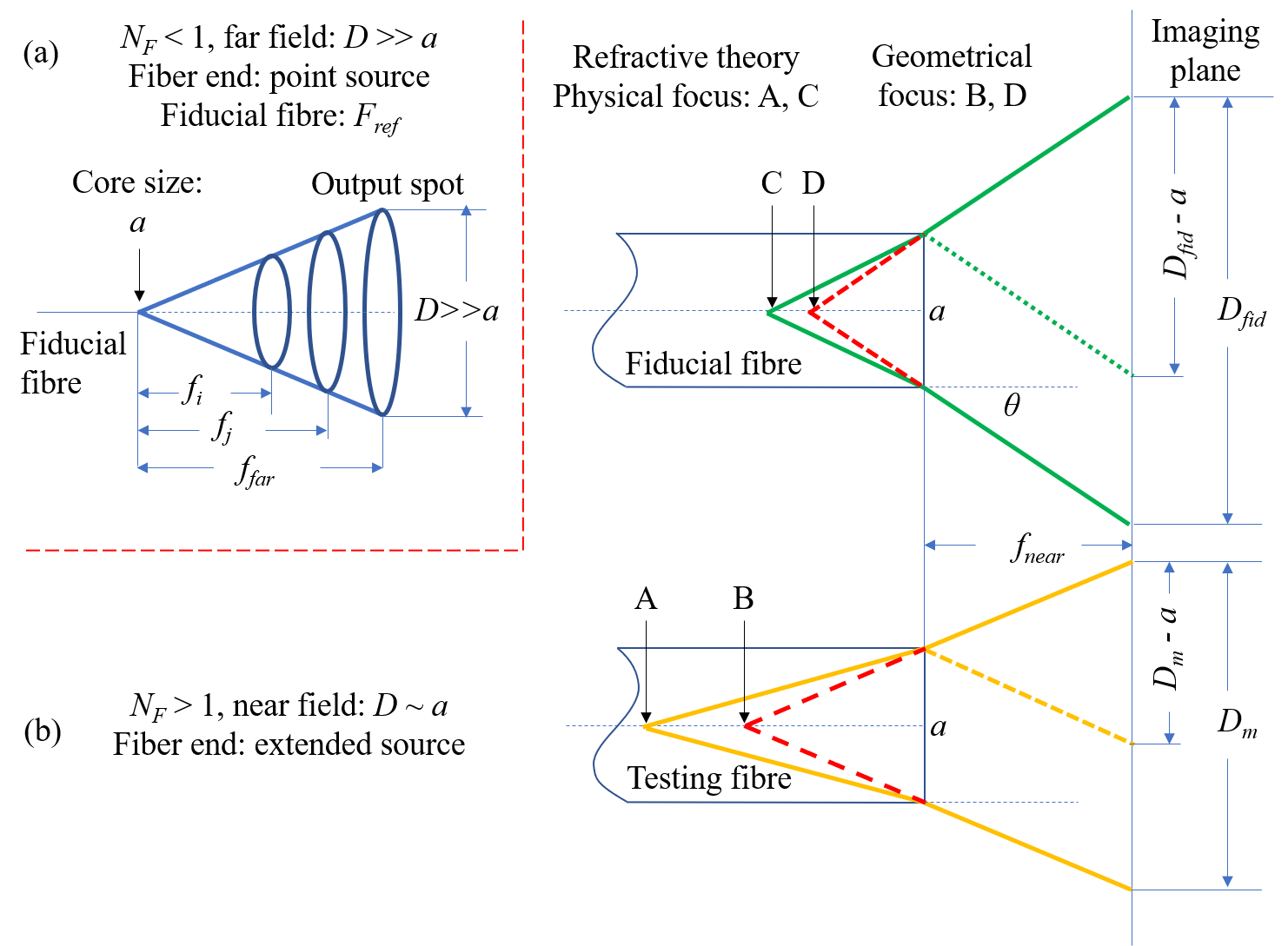}
  \caption{The figure illustrates the difference in FRD measurement between the near and far field for different Fresnel number conditions. In the case where the Fresnel number $N_F<1$ (a), the far field measurement assumes the fibre to be a point source, and the output focal ratio of the reference focal ratio $F_{ref}$ of the fiducial fibre is determined using the common imaging method. However, for conditions where $N_F>1$ (b), the fibre end behaves more like an extended source, altering the geometric relationship between the fiducial fibre and the testing fibre. Consequently, before applying the quasi-near field method to measure FRD for FASOT-IFU, it is crucial to perform error analysis. As an extended source, the physical focus and geometrical focus do not coincide, making it more sensitive to variations in the angle of the fibre end face with respect to the imaging plane. This sensitivity must be taken into account when considering the uncertainties in the FRD measurement.
  }\label{fig.nearandfarfield}
\end{figure}

In the far field, the Fresnel number is smaller than 1, and the output spot is larger than the fibre core, which can be treated as a point source. Output spots in several positions can be recorded and the reference focal ratio of the fiducial fibre can be determined by the common imaging method. According to the geometrical relationship, the reference focal ratio is the ray on surface of the light cone as shown by the orange solid line in Fig.~\ref{fig.nearandfarfield}. In the quasi-near field, the Fresnel number is larger than 1, and the output spot has the similar size of the fibre core. Therefore, the fibre core should be excluded as shown by orange dashed line in Fig.~\ref{fig.nearandfarfield}, and the theoretical focal ratio in geometry can be derived using equation~(\ref{eq.focalratioinnearfield}):
\begin{equation}\label{eq.focalratioinnearfield}
  \frac{{{F_{theo}}}}{{{F_{fid}}}} = \frac{{{{{f_{near}}} \mathord{\left/
 {\vphantom {{{f_{near}}} {\left( {{D_m} - a} \right)}}} \right.
 \kern-\nulldelimiterspace} {\left( {{D_m} - a} \right)}}}}{{{{{f_{near}}} \mathord{\left/
 {\vphantom {{{f_{near}}} {\left( {{D_{fid}} - a} \right)}}} \right.
 \kern-\nulldelimiterspace} {\left( {{D_{fid}} - a} \right)}}}} = \frac{{{D_{fid}} - a}}{{{D_m} - a}}
\end{equation}

The relative error of the focal ratio of the testing fibre between the measured and theoretical values can be calculated using equation~(\ref{eq.relativef}):
\begin{equation}\label{eq.relativef}
  \nabla F = \frac{{\left| {{F_m} - {F_{theo}}} \right|}}{{{F_{theo}}}} = \frac{\varepsilon }{{\frac{{{D_{fid}}}}{a} - 1}} = \frac{\varepsilon }{{\frac{a}{{{N_F}{F_{theo}}\lambda }} - 1}}
\end{equation}
where, $\varepsilon  = \frac{{\left| {{D_m} - {D_{fid}}} \right|}}{{{D_m}}}$ represents the relative difference of the diameters of the output spots of the fiducial fibre and the testing fibres. The relative error $\nabla F$ has the inverse relation with the ratio of $D_{fid}/a$, where a large output spot acquired in a relatively far distance can reduce the relative error and improve the accuracy of the focal ratio. In the focal ratio measurement of the fiducial fibre in the far field, the value of $D_{fid}/a$ is lager than 100, and the relative error $\nabla F$ is less than $10^{-4}$. Thus, the focal ratio of the fiducial fibre can be considered as the same as the theoretical value ($F_{fid} = F_{theo}$), and thus the absolute error in $F$-ratio can be calculated using equation~(\ref{eq.deltaf}):
\begin{equation}\label{eq.deltaf}
  \Delta F = \left| {{F_{theo}} - {F_m}} \right| = \nabla F \cdot {F_{theo}} = \nabla F \cdot {F_{fid}} = \frac{{\varepsilon {F_{fid}}}}{{\frac{a}{{{N_F}{F_{fid}}\lambda }} - 1}}
\end{equation}

Equations~(\ref{eq.relativef}) and (\ref{eq.deltaf}) describe the dependence of the error on the reference focal ratio in the parallel analogy method. In practical applications, the reference focal ratio should not be too large since a large reference focal ratio will increase the absolute error $\Delta F$. However, a small reference focal ratio can also increase the output spot diameter, increasing the chances of overlap of the adjacent spots. Therefore, the reference focal ratio $F_{fid}$ should be optimized to be similar to the value of the testing fibre, which can guarantee the spots are separated and decrease the relative error $\nabla F$. Then, by applying error propagation to equation~(\ref{eq.deltaf}), the tolerance of the uncertainty range of the absolute error is as follows:
\begin{equation}\label{eq.errorrange}
  \Delta {F_{tol}} = \frac{{\varepsilon a{N_F}{F_{fid}}\lambda }}{{{{\left( {a - {N_F}{F_{fid}}\lambda } \right)}^2}}}{\left[ {{{\left( {\frac{{\delta {F_{fid}}}}{{{F_{fid}}}}} \right)}^2} + {{\left( {\frac{{\delta {N_F}}}{{{N_F}}}} \right)}^2}} \right]^{\frac{1}{2}}}
\end{equation}
where, \[\delta {N_F} = {\left[ {{{\left( {\frac{{2{N_F}}}{a}\delta a} \right)}^2} + {{\left( {\frac{{\lambda {N_F}^2}}{{{a^2}}}\delta L} \right)}^2}} \right]^{\frac{1}{2}}}.\]
Equation~(\ref{eq.errorrange}) gives the absolute error in focal ratio of parallel analogy method of the FRD measurement in different conditions.

\subsection{FRD results for FASOT-IFU}

FASOT has two IFUs, each containing both science fibres and a fiducial fibre psitioned in the corner surrounding the microlens array. In the experiment, the input light remains constant for both the fiducial fibre and the science fibres. Initially, the output focal ratio of the fiducial fibre is measured independently as a reference. This is achieved by obstructing the science fibres with a diaphragm and allowing the light to only couple into the fiducial fibre. The output focal ratio of the fiducial fibre is then determined using the common imaging method.

Subsequently, the diaphragm is removed, and all the fibres, including the science fibres, are illuminated. The FRD performance is measured using the parallel analogy method. Due to the close arrangement of the fibres in the V-groove, the distance between the fibre end and the imaging plane is constrained to a range of 500-700\,nm to avoid the overlap of the output spots. With this limitation, the error in the FRD measurement using the parallel analogy method under these conditions can be estimated according equation~(\ref{eq.errorrange}), as illustrated in Fig.~\ref{fig.ferrofFASOTIFU}.

\begin{figure}
  \centering
  \includegraphics[width=\columnwidth]{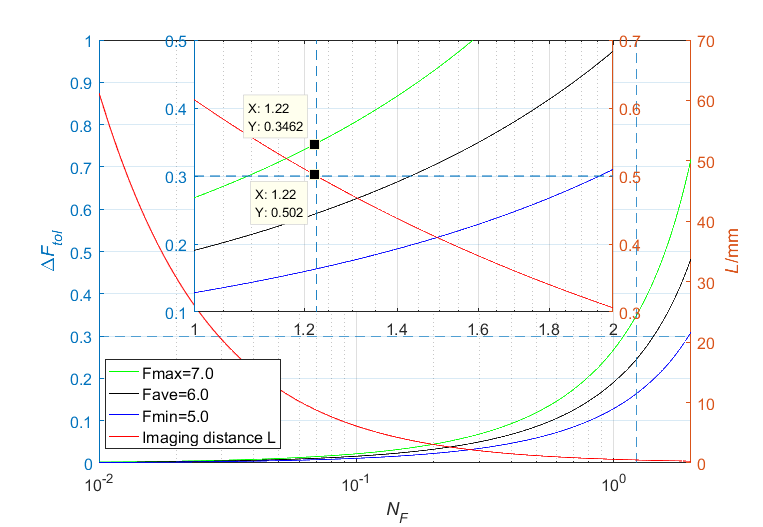}
  \caption{The figure displays the absolute error in the $F$-ratio of the quasi-near field method for FRD measurement in FASOT-IFU. The error analysis considers an output focal ratio range of 5.0-7.0, which closely matches the measured results. The results show that the error in the $F$-ratio increases with higher values of $N_F$ (indicating the imaging plane is closer to the fibre end), consistent with the error analysis presented in Fig.~\ref{fig.nearandfarfield}. For FRD measurement in FASOT-IFU with $N_F=1.225$, the error in the $F$-ratio can be constrained to be less than 0.35 (point A) while maintaining a lower limit on the imaging distance of 500\,$\umu$m (point B). In practical measurements, since most fibres have an output focal ratio larger than 5.0, the imaging distance can be extended beyond 600\,$\umu$m, allowing for further reduction in the error in the $F$-ratio.
  }\label{fig.ferrofFASOTIFU}
\end{figure}

Based on the error distribution analysis, the error of the parallel analogy method for FRD measurement in the case of the FASOT-IFU is found to be smaller than 0.35 within the focal ratio range of 5.0-7.0. This analysis assumes a Fresnel number of $N_F = 1.22$ for an imaging distance of 500 nm. It is important to note that for longer imaging distances, the Fresnel number decreases, resulting in even smaller errors. Therefore, with increased imaging distances, the error in the FRD measurement using the parallel analogy method would be significantly reduced.


The FRD measurement process for the FASOT-IFU involves several steps. Firstly, the common imaging method is used to record the output spots in the far field of the fiducial fibre, resulting in a reference focal ratio of 6.02 with the correction of the magnification effect. The output spot in the quasi-near field is then imaged to obtain the reference diameter $D_{fid}$ (Fig.\ref{fig.farfieldspots}). Next, the entire microlens array is illuminated, and the spots in the quasi-near field of the testing fibres are recorded. Finally, the output focal ratio of the testing fibres is calculated using equations~(\ref{eq.focalratiomodify}) and (\ref{eq.pam}), as shown in Fig.~\ref{fig.frdifu}.

\begin{figure}
  \centering
  \includegraphics[width=\columnwidth]{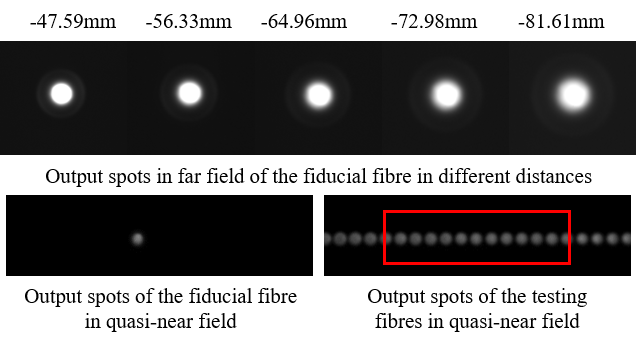}
  \caption{The figure presents the far field spots of the fiducial fibre and the quasi-near field spots of both the fiducial and testing fibres. In the upper part of the figure, the FRD determination of the fiducial fibre is performed using the common imaging method in the far field. The imaging planes are positioned at different distances, represented as relative values read from the scaleplate of the displacement stage. It is important to note that these distances do not correspond to the true positions with respect to the fibre end. The reference focal ratio of the fiducial fibre is evaluated through linear fitting, as depicted in Fig.~\ref{fig.frdexampled}. In the lower part of the figure, the diameters of the quasi-near field spots of the testing fibres (considering only the spots within the red box, which exhibit minimal aberrance) are compared with the diameter ($D_{fid}$) of the fiducial fibre, also acquired in the quasi-near field. The output focal ratios are then calculated based on these measurements.
  }\label{fig.farfieldspots}
\end{figure}

\begin{figure}
  \centering
  \includegraphics[width=\columnwidth]{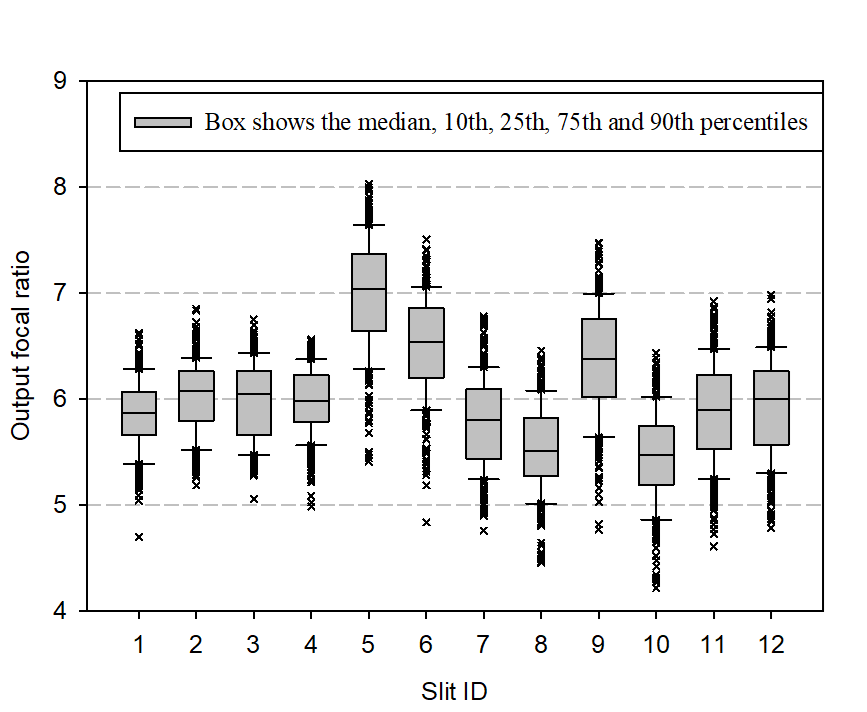}
  \caption{The FRD results of the FASOT-IFU show that each slit, consisting of 336 fibres, exhibits a focal ratio that primarily falls within the range of 5.0-7.0.}\label{fig.frdifu}
\end{figure}

The results indicate that the output focal ratio of the FASOT-IFU primarily falls within the range of 5.0-7.0. Only 1.6 per cent (129 out of 8064 fibres) of the fibres exhibit a significant FRD, with an output focal ratio smaller than 5.0, which might be caused by the stress effect of the glue in the V-groove or the polishing process. Other imperfections that cause serious FRD on the FASOT-IFU should be further studied, as well as the throughput, cross-talk, spectral transmission, etc, will be in the next paper of Sun et al (in preparation). Overall, the FRD performance of the FASOT-IFU meets the design requirements satisfactorily.

\section{Discussion and conclusions}

Building a telescope, especially a large one, is a challenging task that requires significant time and resources. As more instruments are integrated with fibres, it becomes crucial to have precise knowledge of fibre properties for designing instruments that meet the requirements of different scientific goals. The main properties of fibres in astronomy that are of interest to researchers include throughput, FRD, and scrambling, especially for high spectral resolution in fibre-fed spectrographs.

The modified method proposed in this paper offers a convenient way to acquire these fibre parameters. The imaging system can record the output spots in the far field, and by adjusting the magnification ratio and the observing plane, it can also image the power distribution on the fibre end in the near field. The spots can then be used to evaluate the throughput and transmission efficiency by analyzing the grey values of the images, which represent the power transfer. The spots in the near and far fields can be used to investigate FRD and scrambling performance by analyzing size changes and power distributions.

However, the quasi-near field model has some limitations that affect the measurement uncertainties. Firstly, the measurement is sensitive to the aberrations and magnification ratio of the imaging system. Generally, an imaging system with a small magnification ratio provides better imaging quality with minimal aberrations. It is important to note that only the images in the central area of the imaging plane are valid for accurate FRD determination, as aberrations can distort spots in the outer regions, as shown in Fig.~\ref{fig.aberrance}. Moreover, a large magnification ratio reduces the energy density of the spots on the camera sensor, leading to lower image quality and decreased signal-to-noise ratio (SNR).

Secondly, to control the uncertainties in FRD measurements, the parallel analogy method in the quasi-near field model is more accurate when the relative difference between the reference focal ratio and the testing fibres is small. Therefore, it is recommended to preestimate an empirical focal ratio close to the testing fibres as a prior reference.

Lastly, operating in the quasi-near field is challenging and demands precise and mechanically robust structures. The test bench design and setup require high precision and strength to ensure accurate measurements, further adding to the difficulty of the process.


Indeed, the modified method presents a valuable approach for characterizing fibre properties and has the potential to help improve the design and enhance the performance of instruments in terms of throughput, FRD, and scrambling. The successful application of FRD measurement for FASOT-IFU demonstrates the feasibility of the quasi-near field model and the parallel analogy method. This method enables rapid FRD measurement with an acceptable accuracy level of <0.35 in terms of the $F$-ratio for a multiplexed fibre system. It significantly improves measurement efficiency and reduces manual labor.

The FRD results obtained for FASOT-IFU highlight the presence of serious FRD in some fibres. To address this issue, further investigation is necessary to identify the primary cause, which could potentially be related to factors such as polishing or stress effects. It is crucial to improve the assembling technology to achieve better FRD performance in future designs.

Overall, the modified method in the quasi-near field offers a practical and efficient solution for FRD measurement, enabling researchers to optimize fibre performance and enhance the design and construction of fibre-integrated instruments.

\section*{Acknowledgements}

This research was supported by the National Natural Science Foundation of China (NSFC) (12103015); Joint Research Fund in Astronomy through cooperative agreement between the National Natural Science Foundation of China (NSFC) and Chinese Academy of Sciences (CAS) (U1631239, U2031130); Fundamental Research Funds for the Central Universities to the Harbin Engineering University.


\section*{Data Availability}

The data underlying the research results described in the article will be shared by reasonable request to the corresponding author.




\bibliographystyle{mnras}
\bibliography{allrefs} 




%
%


\bsp	
\label{lastpage}
\end{document}